\documentclass[preprint,english,showpacs,12pt,floatfix]{revtex4-1}
\usepackage{babel,amsmath,amssymb,dcolumn}
\usepackage{hyperref}
\usepackage{amsfonts}
\usepackage{amssymb}
\usepackage{graphicx}
\usepackage{graphics}
\usepackage{latexsym}
\usepackage{dcolumn}
\usepackage{epsfig}
\usepackage{subfigure}
\usepackage{array}
\usepackage{multirow}
\usepackage[retainorgcmds]{IEEEtrantools}
\allowdisplaybreaks[1]
\begin{document}

\newcommand{\vp}{\varphi}
\newcommand{\nn}{\nonumber\\}
\newcommand{\beq}{\begin{equation}}
\newcommand{\eeq}{\end{equation}}
\newcommand{\bed}{\begin{displaymath}}
\newcommand{\eed}{\end{displaymath}}
\def\bea{\begin{eqnarray}}
\def\eea{\end{eqnarray}}
\newcommand{\veps}{\varepsilon}
\newcommand{\nablasl}{{\slash \negthinspace \negthinspace \negthinspace \negthinspace  \nabla}}
\newcommand{\om}{\omega}

\newcommand{\Dsl}{{\slash \negthinspace \negthinspace \negthinspace \negthinspace  D}}
\newcommand{\tDsl}{{\tilde \Dsl}}
\newcommand{\tnablasl}{{\tilde \nablasl}}
\title{Quantization of massive spinor fields around global monopoles}

\author{Owen Pavel Fern\'{a}ndez Piedra$^{1, 2}$ \\
%EndAName
\textit{$^1$ Departamento de F\'isica, Divisi\'on de Ciencias e Ingenier\'ias, Universidad de Guanajuato, Campus Le\'on, Loma del Bosque N0. 103, Col. Lomas del Campestre, CP 37150, Le\'on, Guanajuato, M\'exico.}\\
\textit{$^2$Grupo de Estudios Avanzados, Universidad de Cienfuegos, Carretera a Rodas, Cuatro Caminos, s/n. Cienfuegos, Cuba.}}
\email{opavelfp2006@gmail.com}

\begin{abstract}
The renormalized quantum stress energy tensor $\left<T_{\mu}^{\nu}\right>_{ren}$ for a massive spinor field around global monopoles is constructed within the framework of Schwinger-DeWitt approximation, valid whenever the Compton length of the quantum field is much less than the characteristic radius of the curvature of the background geometry. The results obtained shows that the quantum massive spinor field in the global monopole spacetime violates all the pointwise energy conditions.
\end{abstract}
\pacs{04.62.+v,04.70.-s}
\date{\today}
\maketitle

%%%%%%%%%%%%%%%%%%%%%%%%%%%%%%%%%%%%%%
\section{Introduction}
In Quantum Field theory in curved spacetime the most important role is played by the renormalized quantum stress energy tensor \(\langle T_{\mu}^{\nu}\rangle\) of the quantum field, which not only gives information on the energy, pressures and general features of the quantum field itelf, but used as a source in the so called semiclassical Einstein's equations allow to find te changes in the background geometry caused by the quantization of matter fields \cite{DeWitt,birrel,york,lousto-sanchez}. Due to this fact, an imminent need is to have analytical expressions of the renormalized stress tensor.

However, the problem of determine exactly \(\langle T_{\mu}^{\nu}\rangle\) for a generic spacetime is a very difficult, and we need to resort to approximations. There is a vast literature regarding this problem, with works that range from analytical approaches to purely numerical calculations \cite{candelas,koffman,frolov-zelnikov,avramidi,AHS,matyjasek1,matyjasek,berej-matyjasek,owen1,owen2,Folacci}.

One of the most useful approaches, valid for massive fields, is the Schwinger-DeWitt proper time approach, that allow to obtain analytical expressions for the renormalized one-loop effective action for the quantum field, as an expansion in the square of the inverse mass of it. From the effective action, the quantum stress energy tensor can be determined by functional differentiation with respect to the metric. This approach can be used whenever the Compton's wavelenght of the field is less than the characteristic radius of curvature \cite{DeWitt,frolov-zelnikov,avramidi,AHS,matyjasek1,matyjasek,berej-matyjasek,owen1,owen2,Folacci}.

In previous paper \cite{owenmonopole2} we study the quantization of a massive scalar field with arbitrary coupling to the gravitational field of a point-like global monopole, using the Schwinger-DeWitt technique. We obtained very simple analytical results for the components of the quantum stress tensor, and use this results to show that the scalar field violates all the pointwise energy conditions in this spacetime. Global monopoles are interesting heavy objects as they appeared in the early universe as a result of a phase transition of a self-coupled scalar field triplet whose original global $O(3)$ symmetry is spontaneously broken to $U(1)$, being the scalar field the order parameter which is nonzero outside the monopole's core, where the main part of the monopole's energy is concentrated \cite{barriola-vilenkin}. In this paper we continue the investigation of vacuum polarization effects in the spacetime of a poitlike global monopole, using the Schwinger-DeWitt approach to determine the renormalized stress tensor for a quantum massive spinor field in this gravitational background.

Previous works concerning the quantization of fields around global monopole systems includes the analysis of massless scalar fields \cite{hiscock,mazzitely-lousto,bezerra1}, and the determination of the quantum stress energy tensor for a massless spinor field \cite{bezerra2,bezerra3}.

The most simple model which gives rise to global monopoles was constructed by Barriola and Vilenkin in \cite{barriola-vilenkin}. The line element that describes the spacetime geometry around the global monopole is given by
\begin{equation}
ds^2 = -\left(1 - 8\pi \eta^2 - 2 M/r\right)dt^2 + \left(1 - 8\pi \eta^2 - 2 M/r\right)^{-1}dr^2 + r^2(d\theta^2 + \sin^2\theta
d\varphi^2)\ ,
\end{equation}
where $M$ is the mass parameter and $\eta$ is of order $10^{16}Gev$ for a typical grand unified theory. Neglecting the mass term we obtain the line element that describes the geometry around a pointlike global monopole:
\begin{equation}
ds^2 = -\alpha^2 dt^2 + dr^2/\alpha^2 + r^2(d\theta^2 +
\sin^2\theta d\varphi^2)\ ,
\label{metric}
\end{equation}
where we define the parameter $\alpha$ according to the expression $\alpha^2 = 1 - 8\pi \eta^2$.

If we re-scale in the above solution the time and radial variables using $\tau=\alpha t$ and $\rho=\frac{r}{\alpha}$ we arrive to the line element
\begin{equation}
ds^2 = -d\tau^2 + d\rho^2 + \alpha^{2}\rho^2(d\theta^2 +
\sin^2\theta d\varphi^2)\ ,
\label{metric2}
\end{equation}
which shows that, as the value of $\eta$ in field theory predicts a value for $\alpha< 1$, this spacetime is characterized by a solid angle deficit, defined as the difference between the solid angle
in the flat spacetime $4\pi$ and the solid angle in the global monopole spacetime $4\pi\alpha^2$.

In the following consider a single massive neutral spinor field with mass $\mu$ in the gravitational background of a pointlike global monopole in four dimensions
The action for the system is:
\begin{equation}\label{}
    S=\frac{1}{16\pi}\int d^{4}x\sqrt{-g} \ R+\frac{i}{2}\int d^{4}x\sqrt{-g}\widetilde{\phi}\left[\Gamma^{\mu}\nabla_{\mu}\phi+ \mu\phi\right]\label{action}
\end{equation}
where \(\phi\) provides a spin representation of the vierbein group and \(\widetilde{\phi}=\phi^{*}\Gamma\), where * indicates the operation of transposition. \(\Gamma\) and \(\Gamma^{\mu}\) are the curved space Dirac matrices with satisfy the usual relations \(\left[\Gamma^{\mu},\Gamma^{\nu}\right]_{+}=2g^{\mu\nu}\widehat{I}\),
being \(\widehat{I}\) the \(4\times4\) unit matrix. The covariant derivative of any spinor \(\chi\) obey the conmutation
relations \cite{deWitt2}:
\begin{equation}\label{}
    \left[\nabla_{\mu},\nabla_{\nu}\right]\chi=\frac{1}{2}\mathfrak{F}_{[\alpha,\beta]} \ R^{\alpha\beta}_{\ \ \ \mu\nu}
\end{equation}
\begin{equation}\label{}
    \left[\nabla_{\nu},\nabla_{\sigma}\right]\nabla_{\mu}\chi=\frac{1}{2}\mathfrak{F}_{[\alpha,\beta]} \ R^{\alpha\beta}_{\ \ \
    \mu\sigma}\nabla_{\mu}\chi \ + \ R_{\mu \ \ \nu\sigma}^{\ \ \rho} \ \nabla_{\rho}\chi
\end{equation}
\begin{equation}\label{}
   \left[\nabla_{\sigma},\nabla_{\tau}\right]\nabla_{\nu}\nabla_{\mu}\chi=\frac{1}{2}\mathfrak{F}_{[\alpha,\beta]} \ R^{\alpha\beta}_{\ \ \
    \sigma\tau}\nabla_{\nu}\nabla_{\mu}\chi \ + \ R_{\mu \ \ \sigma\tau}^{\ \ \rho} \ \nabla_{\nu}\nabla_{\rho}\chi \ + \ R_{\nu \ \ \sigma\tau}^{\ \ \rho} \ \nabla_{\rho}\nabla_{\mu}\chi
\end{equation}
and so forth, where \(\left[\ \ , \ \
\right]\) is the commutator bracket, $\mathfrak{F}_{\left[\alpha,\beta\right]}=\frac{1}{4}\left[\Gamma_{\alpha},\Gamma_{\beta}\right]$ are the generators of the vierbein group and $R^{\alpha\beta}_{\ \ \ \mu\nu}=h^{\alpha}_{\ \sigma} \ h^{\beta}_{\ \tau} \ R^{\sigma\tau}_{\ \ \ \mu\nu}$, being \(h^{\alpha}_{\ \beta}\) the vierbein which satisfies \(h_{\alpha\mu} \ h^{\alpha}_{\ \nu}=g_{\mu\nu}\). The covariant derivatives of \(\Gamma\), \(\Gamma^{\mu}\) and \(\mathfrak{F}_{\left[\alpha,\beta\right]}\) vanishes.

From the action (\ref{action}) we obtain the equation of motion for the massive spinor field $\phi$:
\begin{equation}\label{}
    \left(\Gamma^{\mu}\nabla_{\mu}+m\right)\phi=0 \label{diraceqn1}
\end{equation}

In the following, we apply the Schwinger-DeWitt technique to obtain the one-loop effective action for the massive spinor fields that obeys (\ref{diraceqn1}). The Schwinger-DeWitt technique is directly applicable to "minimal" second order differential operators ( acting on the super-field \(\phi^{A}\)) that have the
general form:
\begin{equation}\label{}
     \hat{D}=\Box-\mu^{2}+Q \label{minimalop}
\end{equation}
where \(\Box\,=\,g^{\mu\nu}\nabla_{\mu}\nabla_{\nu}\) is the covariant D'Alembert operator, \(\nabla_{\mu}\) is the covariant derivative defined by means of some background connection \(\mathfrak{C}_{\mu}\left(x\right)\) as $\nabla_{\mu}\phi^{A}=\partial_{\mu}\phi^{A}+\mathfrak{C}^{A}_{\ B\mu}\phi^{B}$, \(\mu\) is the mass field and \(Q^{A}(x)\) is an arbitrary matrix representing the potential.

We don't need the explicit form of the background affine connection \(\mathfrak{C}^{A}_{\ B\mu}(x)\) that defines the covariant
derivative above, but it is only necessary to know the commutator of covariant derivatives that defines
curvature $\left[\nabla_{\alpha},\nabla_{\beta}\right]\phi=\mathfrak{R}_{\alpha\beta}\phi $, with $\mathfrak{R}_{\alpha\beta}=\partial_{\alpha}\mathfrak{C}_{\beta}-\partial_{\beta}\mathfrak{C}_{\alpha}+\left[\mathfrak{C}_{\alpha},\mathfrak{C}_{\beta}\right]$
In the spinor field the curvature has the form $\mathfrak{R}_{\alpha\beta}=\gamma^{\sigma}\gamma^{\tau}R_{\sigma\tau\alpha\beta}$.

As is usual in Quantum Field Theory the effective action of the quantum field \(\phi\) can be given as a perturbation
expansion in the number of loops $ W \left(\Phi\right)= S \left(\Phi\right)+\sum_{k\geq1} \ W^{(k)}\left(\Phi\right)$ where \(S\left(\Phi\right)\) is the classical action of the free field. At one-loop we have:
\begin{equation}\label{}
    W^{(1)}=-\frac{i}{2}\ln\left(s\det\hat{D}\right)
\end{equation}
where \(s\det\hat{F}=\exp(str\ln\hat{D})\) is the functional Berezin superdeterminant of the operator \(\hat{D}\), and $str \hat{F}=\left(-1\right)^{i}D^{i}_{i}=\int d^{4}x\left(-1\right)^{A}{D}^{A}_{A}(x)$ is the functional supertrace \cite{avramidi}.

Now using the Schwinger-DeWitt representation for the Green's function of the operator (\ref{minimalop}), we can obtain the renormalized one-loop effective action of the quantum field \(\phi\):
\begin{equation}\label{}
         W^{(1)}_ {ren}\,=\,{1\over 32\pi^{2}\,}\,\int d^{4}x \sqrt{-g}\,\sum_{k=3}^{N}\frac{\left(k-3\right)!}{\mu^{2(k-2)}}\left[a_{k}\right]
\end{equation}

The quantities $[a_{k}]= \lim_{x'\rightarrow x}\,a_{k}(x,x')$, are the coincident limits of the Hadamard-DeWitt coefficients, whose complexity rapidly increases with \(k\). The results for this coefficients up to order $k=4$ can be find, for example, in reference \cite{avramidi}. The first three coefficients $[a_{0}],\,[a_{1}],\,{\rm and}\,[a_{2}], $ contribute to the divergent part of the action and can be absorbed in the classical gravitational action by renormalization of the bare gravitational and cosmological constants.

As the differential operator \(\hat{A}=\gamma^{\mu}\nabla_{\mu}+m\) in (\ref{diraceqn1}) is not of the appropiate form
(\ref{minimalop}), to apply the Schwinger-DeWitt technique we introduces a new spinor variable \(\zeta\) connected with \(\phi\) by the relation
\(\phi=\Gamma^{\sigma}\nabla_{\sigma}\zeta-m\zeta\) so that (\ref{diraceqn1}) take the form:
\begin{equation}\label{}
     \Gamma^{\mu}\Gamma^{\nu}\nabla_{\mu}\nabla_{\nu}\zeta-m^{2}\zeta=0
\end{equation}
Now using the properties of Dirac matrices and the form of the spinor curvature $\mathfrak{R}_{\alpha\beta}$ we can establish the identity
$\Gamma^{\mu}\Gamma^{\nu}\nabla_{\mu}\nabla_{\nu}=\hat{I}\left(\Box-\frac{1}{4}R\right)$ so that equation (\ref{diraceqn1}) takes the desired form:
\begin{equation}\label{}
     \left(\Box-\frac{1}{4}R-\mu^{2}\right)\zeta=0 \label{diraceqn2}
\end{equation}
with the potential matrix given by $Q=-\frac{1}{4}R \hat{I}$.

Using the above relations, integration by parts and the elementary properties of the Riemann tensor, we can show that the one-loop effective action can be written, in the basis proposed in the paper \cite{Folacci}, as

\begin{eqnarray} \label{efactionok}
& & W^{(1)}_{\mathrm{ren}}= \frac{1}{192 \pi^2 \mu^2}\int d^4
x\sqrt{-g} \left[-\frac{3}{280}
  \, R\Box R  +\frac{1}{28}\,  R_{\alpha \beta} \Box R^{\alpha \beta} +\frac{1}{864} \, R^3  - \frac{1}{180}\, RR_{\alpha \beta} R^{\alpha \beta}  \right. \nonumber \\
& & \qquad \left.   - \frac{25}{756} \, R_{\alpha \beta} R^{\alpha}_{\phantom{\alpha} \gamma}R^{\beta \gamma} +\frac{47}{1260}
\, R_{\alpha \beta}R_{\gamma \delta}R^{\alpha \gamma \beta \delta}- \frac{7}{1440} \, R_{\alpha \beta}R^\alpha_{\phantom{\alpha} \gamma \delta \epsilon} R^{\beta \gamma \delta \epsilon} \right. \nonumber \\
& &  \qquad \left.
       +\frac{19}{1260}\, RR_{\alpha \beta \gamma \delta} R^{\alpha \beta \gamma \delta}+ \frac{29}{7560} \, R_{\alpha \beta \gamma \delta}R^{\alpha \beta \sigma \rho}
R^{\gamma \delta}_{\phantom{\gamma \delta} \sigma \rho} - \frac{1}{108}\, R_{\alpha \gamma \beta \delta} R^{\alpha \phantom{\sigma}
\beta}_{\phantom{\alpha} \sigma \phantom{\beta} \rho} R^{\gamma \sigma \delta \rho} \right].
\end{eqnarray}

The renormalized quantum stress energy tensor for the massive spinor field in a generic spacetime background can be determined from (\ref{efactionok}) by functional differentiation with respect to the metric tensor:
\begin{eqnarray}\label{SET}
&&  \langle  ~T_{\mu\nu
} ~  \rangle_{\mathrm{ren}}=\frac{2}{ \sqrt{-g}} \frac{\delta
W^{(1)}_{\mathrm{ren}}} {\delta
g_{\mu\nu }}= \frac{1}{96 \pi^2 \mu^2}\left[\frac{1}{70} \, (\Box R)_{;\mu\nu}
+ \frac{1}{40}\, R \Box R_{\mu \nu} + \frac{23}{840}\,R_{;\alpha (\mu} R^{\alpha}_{\phantom{\alpha}\nu)}-\frac{1}{28} \,  \Box \Box R_{\mu \nu} \right. \nonumber\\
&& \qquad \qquad \left.  -\frac{1}{120}\,  R R_{;\mu \nu}+\frac{1}{120}\, (\Box R) R_{\mu \nu} + \frac{29}{420} \, R_{\alpha (\mu}
\Box
R^\alpha_{\phantom{\alpha} \nu)}- \frac{19}{420} \,R^{\alpha \beta} R_{\alpha \beta;(\mu \nu)}+ \frac{61}{420} \, R^{\alpha \beta} R_{\alpha (\mu ; \nu)\beta}\right.\nonumber \\
&& \qquad \qquad \left.
 -\frac{11}{105} \, R^{\alpha \beta} R_{\mu\nu; \alpha \beta}-\frac{1}{105} \,R^{;\alpha \beta}R_{\alpha \mu \beta \nu}  -\frac{17}{210} \, (\Box R^{\alpha \beta})R_{\alpha \mu \beta \nu}+
 \frac{13}{105} \,R^{\alpha \beta;\gamma}_{\phantom{\alpha \beta;\gamma} (\mu} R_{|\gamma \beta \alpha|
\nu)}   \right. \nonumber \\
&& \qquad \qquad \left. + \frac{16}{105}  \,R^{\alpha \phantom{(\mu}; \beta \gamma}_{\phantom{\alpha } (\mu} R_{|\alpha \beta \gamma|
\nu)} +\frac{1}{210} \,R^{\alpha \beta \gamma \delta} R_{\alpha \beta \gamma \delta ; (\mu \nu)
}+\frac{19}{840} \,
R_{;\alpha } R^\alpha_{\phantom{\alpha} (\mu;\nu)}  - \frac{1}{420} \,R_{;\alpha }R_{\mu\nu}^{\phantom{\mu\nu}; \alpha} \right. \nonumber \\
&& \qquad \qquad \left.
- \frac{1}{60} \, R^{\alpha \beta}_{\phantom{\alpha \beta};(\mu}R_{\nu)\alpha;\beta}  -\frac{1}{140} \,
R^\alpha_{\phantom{\alpha} \mu;\beta} R_{\alpha \nu}^{\phantom{\alpha \nu};\beta} +
 \frac{3}{35} \,R^\alpha_{\phantom{\alpha} \mu;\beta} R^\beta_{\phantom{\beta} \nu;\alpha} -\frac{1}{21}\,
R^{\alpha \beta;\gamma} R_{\gamma \beta \alpha (\mu;\nu) } \right. \nonumber \\
&& \qquad \qquad \left.  -\frac{11}{105} \, R^{\alpha \beta;\gamma} R_{\alpha \mu \beta \nu;\gamma}+\frac{1}{420}\,R^{\alpha \beta \gamma \delta}_{\phantom{\alpha \beta \gamma \delta};\mu} R_{\alpha \beta \gamma \delta ; \nu }-\frac{4}{105} \, R^{\alpha \beta \gamma}_{\phantom{\alpha \beta \gamma}\mu;\delta}R_{\alpha \beta \gamma \nu}^{\phantom{\alpha \beta \gamma
\nu};\delta}  -\frac{1}{288}  \, R^2 R_{\mu\nu}\right. \nonumber \\
&& \qquad \qquad \left. -\frac{7}{360}\, R R_{\alpha \mu} R^\alpha_{\phantom{\alpha}\nu}
+\frac{1}{180} \, R^{\alpha \beta}R_{\alpha \beta}R_{\mu\nu} -\frac{1}{252} \, R^{\alpha \beta}R_{\alpha \mu}R_{\beta \nu}+ \frac{13}{1260} \,R^{\alpha \gamma}R^\beta_{\phantom{\beta} r}R_{\alpha \mu \beta \nu}\right. \nonumber \\
&& \qquad \qquad \left.  + \frac{97}{1260} \, R^{\alpha \beta}R^\gamma_{\phantom{\gamma} (\mu}R_{ |\gamma \beta \alpha|  \nu) }
+ \frac{11}{360} \, R R^{\alpha \beta \gamma}_{\phantom{\alpha \beta \gamma}\mu }R_{\alpha \beta \gamma \nu} + \frac{7}{1440}\,R_{\mu\nu}R^{\alpha \beta \gamma \delta} R_{\alpha \beta \gamma \delta }\right. \nonumber \\
&& \qquad \qquad\left. -\frac{73}{1260}\, R^\alpha_{\phantom{\alpha}  (\mu}R^{\beta \gamma \delta}_{\phantom{\beta \gamma \delta} |\alpha|
}R_{ |\beta \gamma \delta| \nu) } +\frac{19}{504} \, R^{\alpha \beta}R^{\gamma \delta}_{\phantom{\gamma \delta} p\mu}R_{\gamma \delta \beta \nu}+ \frac{73}{1260} \, R_{\alpha \beta}R^{\alpha \gamma \beta \delta}R_{\gamma \mu \delta \nu}\right. \nonumber \\
&& \qquad \qquad \left. -\frac{97}{1260} \, R_{\alpha \beta}R^{\alpha \gamma \delta}_{\phantom{\alpha \gamma \delta}\mu}R^\beta_{\phantom{\beta} \gamma \delta \nu}+ \frac{73}{2520} \,R^{\alpha \beta \gamma \delta}R_{\alpha \beta \sigma \mu }R_{\gamma \delta \phantom{\sigma}
\nu}^{\phantom{\gamma \delta} \sigma}\right. \nonumber \\
&& \qquad \qquad \left.
 + \frac{239}{1260} \, R^{\alpha \gamma \beta \delta}R^\sigma_{\phantom{\sigma}
\alpha \beta \mu}R_{\sigma \gamma \delta \nu}  -\frac{73}{2520} \,  R^{\alpha \beta \gamma}_{\phantom{\alpha \beta \gamma} \delta } R_{\alpha \beta \gamma \sigma}R^{\delta
\phantom{\mu} \sigma}_{\phantom{\delta} \mu \phantom{\sigma} \nu} + \frac{7}{720} \, R R^{\alpha \beta}R_{\alpha \mu \beta \nu}\right. \nonumber \\
&&  \qquad \qquad \left.
+ g_{\mu\nu} \left[  \frac{1}{280} \, \Box \Box R - \frac{1}{240} \, R\Box R +\frac{1}{420} \, R_{;\alpha \beta}
R^{\alpha \beta} +\frac{1}{105} \,  R_{\alpha \beta} \Box R^{\alpha \beta}+ \frac{3}{70} \, R_{\alpha \beta ; \gamma \delta}R^{\alpha \gamma \beta \delta} \right. \right.\nonumber \\
&& \qquad \qquad \left. \left. -\frac{1}{672}  \,R_{;\alpha}R^{;\alpha}+\frac{3}{280} \, R_{\alpha \beta;\gamma} R^{\alpha \beta;\gamma} -\frac{1}{280} \, R_{\alpha \beta;\gamma}
R^{\alpha \gamma;\beta} + \frac{1}{168} \,R_{\alpha \beta \gamma \delta;\sigma} R^{\alpha \beta \gamma \delta;\sigma}
     +\frac{1}{1728}  \, R^3  \right.\right. \nonumber \\
&& \qquad \qquad \left.\left. - \frac{1}{360}\, RR_{\alpha \beta} R^{\alpha \beta} - \frac{1}{945} \,R_{\alpha \beta}
R^{\alpha}_{\phantom{\alpha} r}R^{\beta \gamma}  + \frac{1}{315} \, R_{\alpha \beta}R_{\gamma \delta}R^{\alpha \gamma \beta \delta} -\frac{7}{2880}\,
RR_{\alpha \beta \gamma \delta} R^{\alpha \beta \gamma \delta} \right. \right. \nonumber \\
&& \qquad  \qquad \left. \left.
 + \frac{7}{360}\,R_{\alpha \beta}R^\alpha_{\phantom{\alpha} \gamma \delta \sigma} R^{\beta \gamma \delta \sigma } -\frac{61}{15120}\, R_{\alpha \beta \gamma \delta}R^{\alpha \beta \sigma \rho}
R^{\gamma \delta}_{\phantom{\gamma \delta} \sigma \rho} -\frac{43}{1512} \, R_{\alpha \gamma \beta \delta} R^{\alpha \phantom{\sigma}
\beta}_{\phantom{\alpha} \sigma \phantom{\beta} \rho} R^{\gamma \sigma \delta \rho}\right ]\right].
\end{eqnarray}

The above result is a rather complex expression for the renormalized stress energy tensor for a spinor field in the Schwinger-DeWitt approximation This is valid for any spacetime \cite{belokogne}. The information of the massive spinor field is included in the coefficients accompanying each local geometric term constructed from the Riemmann tensor, its covariant derivatives and contractions.

Using (\ref{metric}) in (\ref{SET}) we obtain, for the components of the renormalized stress energy tensor for the massive spinor field in the pointlike monopole spacetime the very simple result
\begin{equation}
\left<T_{t}^{t}\right>_{ren}=\frac{\left(1-\alpha ^2\right) \left(31 \alpha ^4+31 \alpha ^2+10\right)}{40320 \pi ^2 \mu ^2 r^6}
\label{eden}
\end{equation}
\begin{equation}
\left<T_{r}^{r}\right>_{ren}=\left<T_{t}^{t}\right>_{ren}
\label{p1}
\end{equation}
and
\begin{equation}
\left<T_{\theta}^{\theta}\right>_{ren}=\left<T_{\varphi}^{\varphi}\right>_{ren}=-2\left<T_{t}^{t}\right>_{ren}=\frac{\left(\alpha ^2-1\right) \left(31 \alpha ^4+31 \alpha ^2+10\right)}{20160 \pi ^2 \mu ^2 r^6}
\label{p2}
\end{equation}

\begin{figure}[t]
%\begin{center}
\scalebox{0.69}{\includegraphics{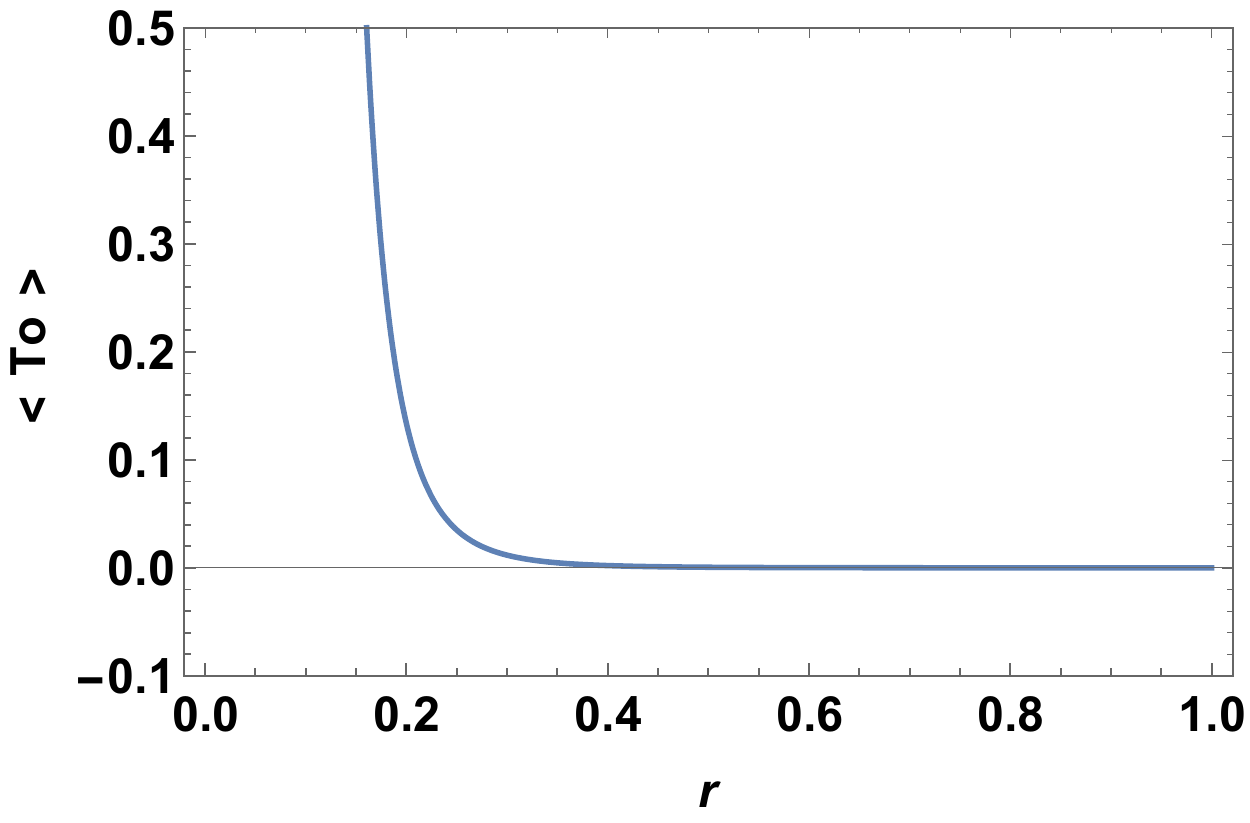}}
\vspace{0.15cm}
\scalebox{0.69}{\includegraphics{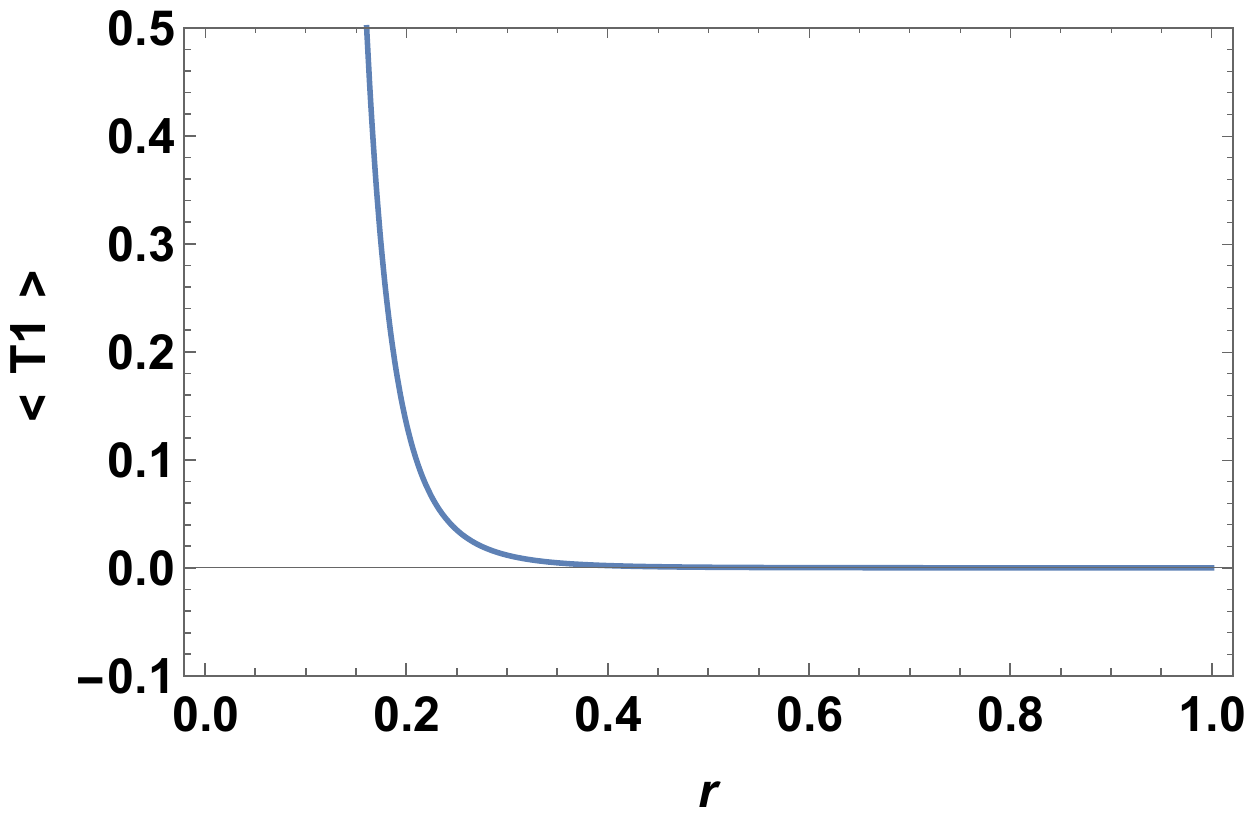}}
\vspace{0.15cm}
\scalebox{0.57}{\includegraphics{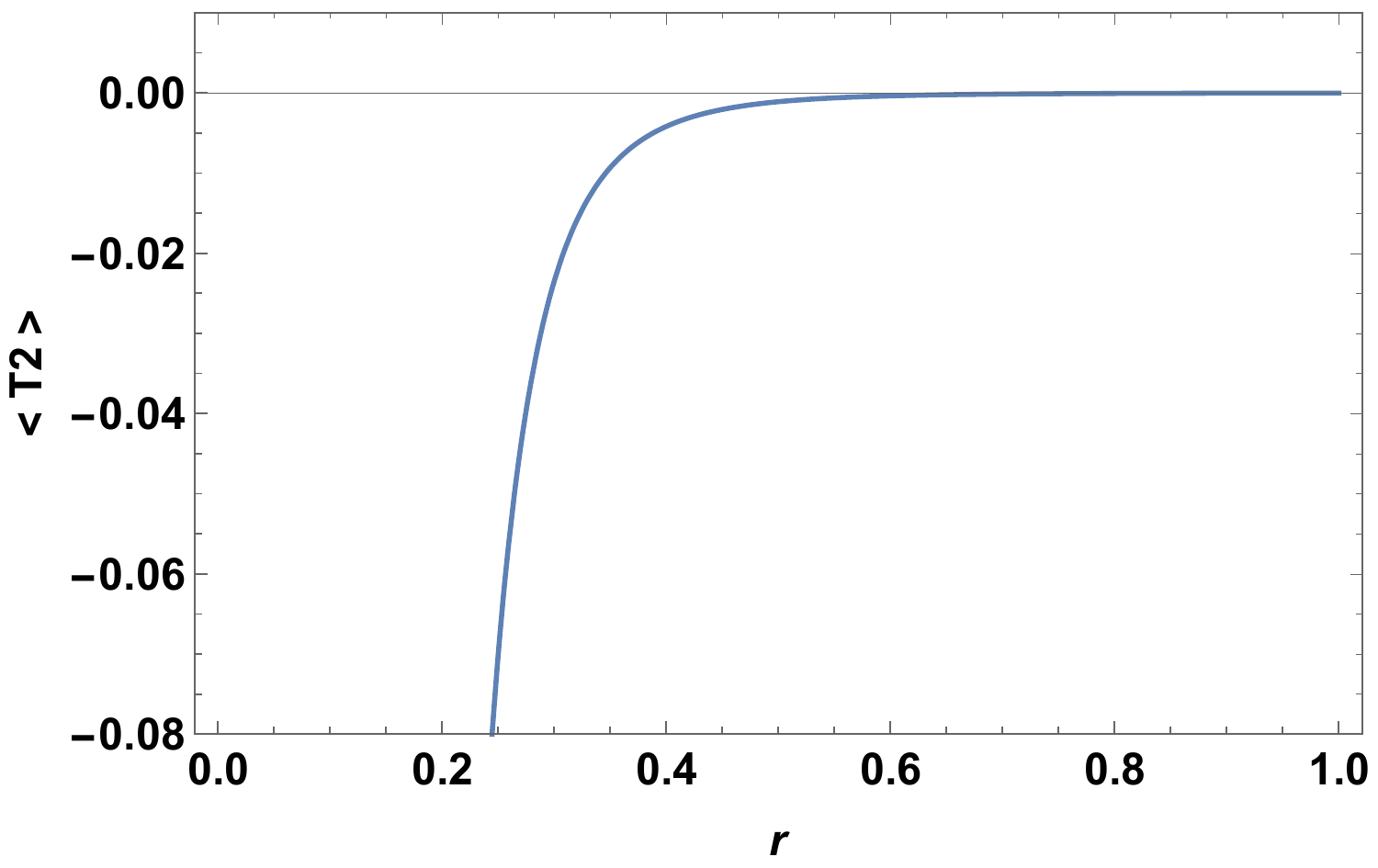}}
%\end{center}
\caption{\textit{Dependance on the distance from monopole's core $r$ of the re-scaled components of the renormalized stress energy tensor $\left<T_{0}\right>=96\pi^{2}\mu^{2}\left <T_{t}^{t}\right >_{ren}$ (top), $\left<T_{1}\right>=96\pi^{2}\mu^{2}\left <T_{r}^{r}\right >_{ren}$ (middle) and $\left<T_{2}\right>=96\pi^{2}\mu^{2}\left <T_{\theta}^{\theta}\right >_{ren}=96\pi^{2}\mu^{2}\left <T_{\varphi}^{\varphi}\right >_{ren}$ (bottom), for a massive spinor field in the pointlike global monopole spacetime. The values of the parameter used in the calculations is $1 - \alpha^2 =10^{-5}$ }.} \label{f1}
\end{figure}

In Figure (\ref{f1}) we show the dependance on the distance from monopole's core $r$ of the re-scaled components of the renormalized stress energy tensor $\left<T_{0}\right>=96\pi^{2}\mu^{2}\left <T_{t}^{t}\right >_{ren}$, $\left<T_{1}\right>=96\pi^{2}\mu^{2}\left <T_{r}^{r}\right >_{ren}$ and $\left<T_{2}\right>=96\pi^{2}\mu^{2}\left <T_{\theta}^{\theta}\right >_{ren}=96\pi^{2}\mu^{2}\left <T_{\varphi}^{\varphi}\right >_{ren}$, for a massive spinor field in the pointlike global monopole spacetime.

As we can observe, the re-scaled time component $\left<T_{0}\right>$  decreases with the increase of the distance from monopole's center, reaching its minimum value equal to zero as $r\rightarrow \infty$. The same behaviour shows $\left<T_{1}\right>$, due to (\ref{p1}). Thus, for all values of the distance $r$, the time and radial components of the renormalized stress energy tensor for the massive spinor field in the pointlike global monopole background are positive.

On the contrary, the re-scaled angular components $\left<T_{2}\right>$, increases with negative values as $r$ increases, until it reach its maximum value equal to zero at large distances.

Defining the energy density and pressures of the quantum field as $\rho=-\left<T_{t}^{t}\right>_{ren}$, $p_{r}=-\left<T_{r}^{r}\right>_{ren}$ and $p_{\theta}=p_{\phi}=p=\left<T_{\theta}^{\theta}\right>_{ren}=\left<T_{\varphi}^{\varphi}\right>_{ren}$, we can use the above results to investigate the fulfillment or not of the pointwise energy conditions for the massive spinor field in the pointlike global monopole background.

The energy conditions are constraints that the components of the stress energy tensor of matter fields do satisfy , and are important in the context of various theorems concerned with positivity of mass, singularities and topological censorship.

There are four pointwise energy conditions \cite{visser,matyjasekEC}. In terms of the energy density and principal pressures of the quantum field, the Null energy condition (NEC) is satisfied when $\rho-p_{r}\geq 0$ and $\rho+p\geq 0$. The Weak energy condition is equivalent to the NEC with the constraint $\rho\geq 0$ added. The Strong energy condition (SEC) is equivalent to NEC with the constraint $\rho-p_{r}+2p\geq 0$, and finally the dominant energy condition (DEC) is equivalent to the restrictions $\rho\geq 0$ and $-\rho\leq p_{i}\leq \rho$.

In terms of the re-scaled energy density and pressures of the quantum massive spinor field $\varrho=96\pi^{2}\mu^{2} \rho$ and $\textrm{p}_{i}=96\pi^{2}\mu^{2} p_{i}$, the results obtained for the renormalized stress energy tensor indicates that, for all values of the distance from monopole's core, we have $\varrho\leq0$, $p_{r}\leq0$, $\varrho-\textrm{p}_{r}=0$, $\varrho+\textrm{p}<0$ and $\varrho-\textrm{p}_{r}+2\textrm{p}<0$. Then, all the pointwise energy conditions are violated by the quantum massive spinor field in the pointlike global monopole spacetime.

The results obtained in this paper can be used to investigate the backreaction of the quantum massive spinor field upon the spacetime around the pointlike global monopole. For this purpose, we can solve perturbatively the semiclassical Einstein's field equations, using the expressions for the quantum stress tensor reported here as a source. Our results on the implementation of this program will be considered in a future report.

%%%%%%%%%%%%%%%%%%%%%%%%%%%%%%
\section*{Acknowledgments}
This work has been supported by TWAS-CONACYT $2017$ fellowship, that allow to the author to do a sabatical leave at Departamento de F\'isica Te\'orica, Divisi\'on de Ciencias e Ingenier\'ias, Universidad de Guanajuato, Campus Le\'on. The author also express his gratitude to Professor Oscar Loaiza Brito, for the support during the research stay at his group, where this work was completed.
%%%%%%%%%%%%%%%%%%%%%%%%%%%%%%%%%%%%

%%%%%%%%%%%%%%%%%%%%%%%%%%%%%%


\begin{thebibliography}{}

%%%%%%%%%%%%%%%%%%%%%%%%%%%%%%%%%%%%%


\bibitem{DeWitt}{B. S. DeWitt, \emph{Phys. Rept} \textbf{53}, 1615 (1984) }.
\bibitem{birrel}{N.D. Birrel  and P. C. Davies, \emph{Quantum Fields in Curved Space }, (Cambridge University Press, Cambridge, 1982) }.
\bibitem{york}{J. W. York, \emph{Phys. Rev.} \textbf{D31}, 775 (1985), D. Page, \emph{Phys. Rev. } \textbf{D25}, 1499 (1982)}.
\bibitem{lousto-sanchez}{C. O. Loust\'{o} and N. Sanchez, \emph{Phys. Lett.} \textbf{212B}, 411 (1988)}.
\bibitem{candelas}{P. Candelas \emph{Phys. Rev.} \textbf{D21}, 2185 (1980)}.
\bibitem{koffman}{L. A. Koffman and V. Sahni \emph{Phys. Lett.} \textbf{B127}, 197 (1983)}.
\bibitem{frolov-zelnikov}{V. P. Frolov and A. I. Zelnikov, \emph{Phys. Lett.} \textbf{115B}, 372 (1982); V. P. Frolov and A. I. Zelnikov, \emph{Phys. Lett.} \textbf{123B}, 197 (1983); V. P. Frolov and A. I. Zelnikov, \emph{Phys. Rev.} \textbf{D29}, 1057
(1984)}.
\bibitem{avramidi}{I. G. Avramidi, \emph{Nucl. Phys.} \textbf{B355}, 712 (1991), I. G. Avramidi, \emph{PhD Thesis}, hep-th/9510140}.
\bibitem{AHS} {P. R. Anderson, W. A. Hisckock and D. A. Samuel, \emph{Phys. Rev.} \textbf{D51}, 4337 (1995)}.
\bibitem{matyjasek1}{J. Matyjasek, \emph{Phys. Rev.} \textbf{D61}, 124019 (2000) }.
\bibitem{matyjasek} {J. Matyjasek, \emph{Phys. Rev.} \textbf{D63}, 084004 (2001) }.
\bibitem{berej-matyjasek} {W. Berej and J. Matyjasek, \emph{Acta. Phys. Pol.} \textbf{B34}, 3957 (2003)}.
\bibitem{owen1}{Owen Pavel Fern\'{a}ndez Piedra and Alejandro Cabo Montes de Oca, \emph{Phys. Rev.} \textbf{D75}, 107501 (2007) }.
\bibitem{owen2}{Owen Pavel Fern\'{a}ndez Piedra and Alejandro Cabo Montes de Oca, \emph{Phys. Rev.} \textbf{D77}, 024044 (2008) }.
\bibitem{Folacci} {Y. Dec\'{a}ninis and A. Folacci, \emph{Class. Quantum Grav.} \textbf{24}, 4777 (2007)}.
\bibitem{owenmonopole2}{O.P. F. Piedra, \emph{arXiv: 1904.09861 [gr-qc]}, (2019)}.
\bibitem{barriola-vilenkin}{M. Barriola and A. Vilenkin  \emph{Phys. Rev. Lett.} \textbf{63}, 341 (1989)}.
\bibitem{hiscock}{W. Hiscock  \emph{Class. Quantum Grav.} \textbf{7}, L235 (1990)}.
\bibitem{mazzitely-lousto}{F. D. Mazziteli and C. Lousto  \emph{Phys. Rev.} \textbf{D43}, 468 (1991)}.
\bibitem{bezerra1}{F C Carvalho and E R Bezerra de Mello  \emph{Class. Quantum Grav.} \textbf{18}, 1637 (2001)}.
\bibitem{bezerra2}{E. R. Bezerra de Mello, V. B. Bezerra, and N. R. Khusnutdinov,  \emph{Phys. Rev. } \textbf{D60}, 063506 (1999)}.
\bibitem{bezerra3}{E. R. Bezerra de Mello,  \emph{Brazilian Journal of Physics} \textbf{31}, 211 (2001)}.
\bibitem{deWitt2}{B. S. DeWitt, \emph{Dynamical Theory of Groups and Fields}, ( Gordon and Breach, New York, 1965)}.
\bibitem{belokogne}{A. Belokogne and A. Folacci \emph{Phys. Rev. } \textbf{D90}, 044045 (2014)}.
\bibitem{visser}{M. Visser \emph{Phys. Rev. } \textbf{D54}, 5116 (1996)}.
\bibitem{matyjasekEC} {J. Matyjasek, \emph{Acta. Phys. Pol.} \textbf{B34}, 3921 (2003)}.


%%%%%%%%%%%%%%%%%%%%%%%%%%%%%%%%%%%%%%%%%%%%%%%%
%%%%%%%%%%%%%%%%%%%%%%%%%%%%%%%%%%%%%%%%%%%%%%%%%%%%%%

%%%%%%%%%%%%%%%%%%%%%%%%%%%%

%%%%%%%%%%%%%%%%%%%%%%%%%%%%%%%%%%%%%%%%%%%%%%%%%%%%%%%%%%%%%%%%%

%%%%%%%%%%%%%%%%%%%%%%%%%%%%%%%%%%%%%%%%%%%%%%%%%%%%%%%%%%%%%%%%%%%%%%%%





%%%%%%%%%%%%%%%%%%%%%%%%%%%%%%%%%%%%%%%%%%%%%%%%%%%%%%%%%%%%%%%%%%%%%%%%%%%%%%%%%%%%%%%%%%%%%%%%%%


%%%%%%%%%%%%%%%%%%%%555
\end{thebibliography}
\end{document}